\documentclass[twocolumn,preprintnumbers,amsmath,amssymb,prb,showpacs]{revtex4}
\usepackage{graphicx}
\usepackage{dcolumn}
\usepackage{bm}
\usepackage{natbib}
\usepackage{txfonts}
\usepackage{fancybox}
\usepackage{braket}
\usepackage{comment}
\usepackage{amsmath}

\begin{document}


\title{Electronic structure of helicoidal graphene:\par
massless Dirac particles on a curved surface with a screw symmetry}

\author{Masataka Watanabe}
\altaffiliation[Now at ]{Kavli Institute for the Physics and Mathematics of the Universe, The University of Tokyo, Kashiwa, Chiba 277-8582, Japan
}
\email{masataka.watanabe@ipmu.jp}
\affiliation{
Department of Physics, The University of Tokyo, Hongo, Tokyo 133-0033, Japan
}

\author{Hisato Komatsu}
\affiliation{
Department of Basic Science, The University of Tokyo, Komaba,
Tokyo 153-8902, Japan
}

\author{Naoto Tsuji}
\altaffiliation[Now at ]{RIKEN Center for Emergent Matter Science (CEMS), Wako 351-0198, Japan}
\author{Hideo Aoki}
\affiliation{
Department of Physics, The University of Tokyo, Hongo, Tokyo 133-0033, Japan{}
}

\begin{abstract}
Massless Dirac particles on the helicoid
are theoretically investigated. 
With its possible 
application being helical graphene, we
explore how the peculiarities of Dirac particles appear on the curved, 
screw-symmetric surface. 
Zweibein is used to 
derive the massless Dirac equation on the 
helicoid and on general curved surfaces.
We show that
bound states of massless Dirac electrons on the helicoid are shown to be absent, and thus
the system is fully characterized by the scattering probabilities and the phase 
shifts. We obtained these quantities from
numerically calculated wavefunctions.
We find the local density of 
states and the phase shifts behave characteristically
 around the axis of the helicoid.
Bound states of massive Dirac electrons on the surface are also shown to be 
absent
as an extension of the above result on massless Dirac electrons.
A comparison with the non-relativistic case is also made.
\end{abstract}
\maketitle

\section{\label{sec:level1}Introduction}

Graphene and massless Dirac fermions
on it have gained persistent attention over the last few decades
as a way of looking at relativity through condensed matter
physics\cite{aoki2013physics}.
While there are a number of studies on graphene,
only a few focus on curved graphene and its electronic stuructures. It 
is, therefore, worth asking the following question: how does the
electronic structure of graphene change if we 
deform it into a curved surface? 
This is interesting in a number of ways. 
(i) Non-relativistic electrons on periodic curved surfaces
have distinctive electronic structures\cite{AokiKoshino} 
and
we expect something even more interesting with Dirac fermions 
due to their spinor wavefunctions 
rather than scalar ones.
(ii) It is also intriguing to see how 
the presence of positive and negative energy states of Dirac fermions 
exerts its effect on curved surfaces. In other words
how the Klein paradox\cite{ma}, i.e.,
potential barriers
cannot reflect massless Dirac particles, takes its form
on curved surfaces. 

These have motivated us to look into Dirac particles on 
the helicoid, one of the simplest curved
surfaces periodic in one direction\cite{Morimoto}.
The surface is also simple in that it is 
a minimal surface, i.e., its mean
curvature vanishes everywhere -- it is the only 
ruled minimal surface other than the plane. 
Possible applications to physical systems include
graphite with a screw dislocation. 
Horn for example reported a spiral growth patterns on natural graphite
as early as 1952\cite{horn1952spiral}, followed by 
more recent studies by Rakovan and Jaszczak\cite{Rakovan}.
There is also a theoretical study by Bird and Preston
of a spiral graphite in terms of Berry's phase\cite{PhysRevLett.61.2863}.

The rest of the paper is organized as follows.
In Section \ref{sec:level2} we
formulate the Dirac equation on general curved surfaces,
and then in Section \ref{threeeeeeee} apply the formalism to the helicoid. 
In Section \ref{parity} we 
show that there are
no bound states of massless Dirac fermions on the helicoid and 
discuss the scattering amplitudes and the phase shifts
of an electron 
off the spiral axis in terms of partial waves.
We go on to discuss the local density of states in Section \ref{five}. 
In Section \ref{endd} we examine the (non-)existence of bound states 
of massive Dirac fermions on the helicoid.
In Appendix \ref{nonre} comparison with
the non-relativistic case is briefly made.

\section{\label{sec:level2}Dirac equation on curved surfaces}
The massless Dirac equation on a flat two-dimensional surface,
\begin{equation}
i\sigma^\mu\partial_\mu\psi=E\psi,
\end{equation}
describes the low-energy 
behaviour of electrons on 
graphene\cite{aoki2013physics}.
Here $\sigma^\mu$ $(\mu=1,\,2)$ are the Pauli matrices,
 $\psi$ is the spinor wavefunction, and
$E$ is the eigenenergy\footnote{The massive case with mass $m$ 
are described by 
adding $im\sigma^3\phi$ term to the left-hand side of the equation.}. 
Here and hereafter repeated indices are implicitely summed over. 
We, following Birrell and Davies\cite{birrell1984quantum},
generalize the Dirac equation to curved surfaces by 
replacing derivatives with covariant derivatives, $D_\mu$,
and by introducing zweibeins, $e^\mu_a$:
\begin{eqnarray}
i\sigma^a e^\mu_aD_\mu\psi=E\psi\quad (a,\,\mu=1,\,2).
\label{eq:d}
\end{eqnarray}
We take a convention in which the metric is given by
$g_{\mu\nu}=e_\mu^ae_\nu^b\eta_{ab}$
with $\eta_{ab} = \mathop{\mathrm{diag}}(+,-,-)$ being 
the Minkowskian metric. The covariant derivatives for spinors are given by
$D_\mu=\partial_\mu+\Gamma_\mu$, where
\begin{equation}\label{gamma}
\Gamma_\mu=\frac{1}{2}S^{ab}e^\nu_ag_{\rho\nu}D_\mu e^\rho_b
\quad\text{ and}
\quad S^{ab}=\frac{1}{4}[\sigma^a,\sigma^b].
\end{equation}
The action of covariant derivatives on zweibeins is
\begin{eqnarray}
D_\nu e^a_\mu = \partial_\nu e^a_\mu +e^b_\mu\omega^a_{b\mu}
-e^a_\lambda\Gamma^\lambda_{\nu\mu}.
\end{eqnarray}
Here $\omega^a_{b\mu}$ are the spin connection
coefficients defined via
\begin{eqnarray}
\label{deff}
d(e^a_\mu dx^\mu)=-\omega^a_b\wedge (e^b_\nu dx^\nu),
\end{eqnarray}
while $\Gamma^\lambda_{\nu\mu}$ are the Christoffel symbols.
We require the spin connection coefficients to be antisymmetric in $a$ and $b$
so as to avoid the arbitrariness in the above definition.

Every surface has global 
isothermal coordinates, i.e.,
every metric on any surfaces can be represented in a diagonal form
upon a suitable coordinate transformation\cite{1955}:
\begin{eqnarray}\label{oha}
g_{ij}(x,y)=
\begin{pmatrix}
g(x,y)^2 & 0 \\
0 & g(x,y)^2
\end{pmatrix}
.
\end{eqnarray} 
The zweibeins and Christoffel symbols then read, in matrix forms,
\begin{eqnarray}
&e^a_\mu=
\begin{pmatrix}
g(x,y) & 0 \\
0 & g(x,y)
\end{pmatrix},\,
e_a^\mu=
\begin{pmatrix}
g(x,y)^{-1} & 0 \\
0 & g(x,y)^{-1}
\end{pmatrix}&, \nonumber \\
&\Gamma^\rho_{\kappa1}=\frac{1}{2}
\begin{pmatrix}
g^{-2}\partial_1(g^2) & g^{-2}\partial_2(g^2) \\
g^{-2}\partial_2(g^2) & -g^{-2}\partial_1(g^2)
\end{pmatrix},& \nonumber \\
&\Gamma^\rho_{\kappa2}=\frac{1}{2}
\begin{pmatrix}
g^{-2}\partial_2(g^2) & -g^{-2}\partial_1(g^2) \\
g^{-2}\partial_1(g^2) & g^{-2}\partial_2(g^2)
\end{pmatrix}.&
\end{eqnarray}
The spin connections are calculated as
\begin{equation}
\begin{cases}
\omega^1_{21}=\displaystyle{\frac{1}{g}\partial_2(g)},\\
\omega^2_{12}=\displaystyle{\frac{1}{g}\partial_2(g)},\\
\omega^a_{\mu b}=0. & \text{(other components)}
\end{cases}
\end{equation}
By substituting these into Eq. (\ref{eq:d}), we arrive at the massless
Dirac equation on a curved surface in terms of isothermal coordinates:
\begin{eqnarray}\label{eq:dc}
i
\begin{pmatrix}
0 & \frac{1}{g}\partial\\
\frac{1}{g}\bar{\partial} & 0
\end{pmatrix}
\begin{pmatrix}
\sqrt{g}\psi_+\\
\sqrt{g}\psi_-
\end{pmatrix}
=E
\begin{pmatrix}
\sqrt{g}\psi_+\\
\sqrt{g}\psi_-
\end{pmatrix},
\label{mad}
\end{eqnarray}
where $\partial\equiv\partial_x-i\partial_y$,
$\bar{\partial}\equiv\partial_x+i\partial_y$ 
and $\psi_+,\, \psi_-$ are two components of the spinor
\footnote{Notice 
we have dropped 1/2 from the ordinary definition of $\partial$
and $\bar{\partial}$.}. 
It is easy to see that Eq. (\ref{eq:dc}) has a chiral 
symmetry\cite{aoki2013physics} 
since
the equation has a solution with eigenenergy $-E$ accompanied with a solution with 
eigenenergy $E$:
\begin{eqnarray}\label{eq:dcinv}
i
\begin{pmatrix}
0 & \frac{1}{g}\partial\\
\frac{1}{g}\bar{\partial} & 0
\end{pmatrix}
\begin{pmatrix}
\sqrt{g}\psi_+\\
-\sqrt{g}\psi_-
\end{pmatrix}
=-E
\begin{pmatrix}
\sqrt{g}\psi_+\\
-\sqrt{g}\psi_-
\end{pmatrix}.
\end{eqnarray}
Moreover, the system has a zero-mode, where $\sqrt{g}\psi_+$
and $\sqrt{g}\psi_-$
are holomorphic and anti-holomorphic, respectively.

\section{Dirac Equation on the Helicoid}
\label{threeeeeeee}
\begin{figure}[htbp]
\begin{center}
\includegraphics[width=0.8\columnwidth]{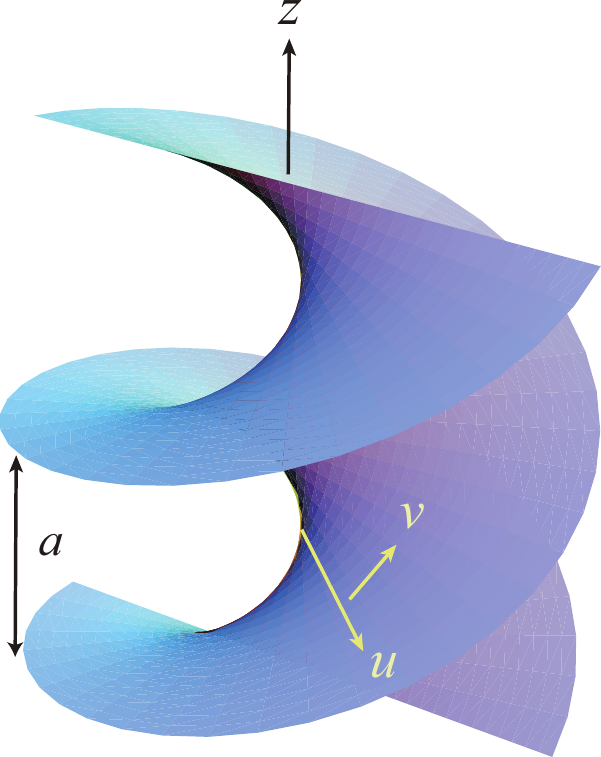}
\end{center}
\caption{(Color online) A helicoid in three-dimensional Euclidean space. 
Shown on the surface are the isothermal coordinates, $(u,v)$.}
\label{rasen}
\end{figure}

The helicoid, shown in Fig. \ref{rasen}, can be parametrized 
in $\mathbb{R}^3\ni (x,y,z)$ as
\begin{equation}
\begin{pmatrix}
x\\ y\\ z\\
\end{pmatrix}
=a
\begin{pmatrix}
\sinh u\cos v\\\sinh u\sin v\\ v
\end{pmatrix}
,\label{3sss}
\end{equation}
where the coordinate $v$ spirals about the axis of the helicoid, while $u$ shoots out 
normal to $v$ from the axis with $a$ being the pitch of the spiral. 
The metric in this parametrization is 
\begin{eqnarray}
g_{ij}=
\begin{pmatrix}
a\cosh^2 u & 0 \\
0 & a\cosh^2 u
\end{pmatrix}
,
\label{metricparam}
\end{eqnarray}
which indicates that $(u,v)$ are indeed isothermal. 

The massless Dirac equation (\ref{eq:dc}) on the helicoid then reads
\begin{equation}
\frac{i}{a\cosh u}\begin{pmatrix}
0 & {\partial_u}-i\partial_v+{\frac{\tanh u}{2}}\\
{\partial_u}+i\partial_v+{\frac{\tanh u}{2}} & 0
\end{pmatrix}
\begin{pmatrix}
\psi _+ \\
\psi _-\end{pmatrix}
=E\begin{pmatrix}
\psi _+\\
\psi _-
\end{pmatrix}.
\end{equation}
Since $v$ is a cyclic coordinate, representing the 
translational symmetry along the axis of the helicoid, 
we can set 
\begin{eqnarray}
\psi_{\pm}(u,v)=\exp(i\ell v)\phi(u)_{\pm},
\end{eqnarray}
in which $\ell$
is physically the ``angular momentum'' along the $z$ axis. 
Since the system is helical and the coordinate
$v$ spirals around the axis of the helicoid,
 $\ell$ does not necessarily have to be integer-valued,
or cannot be quantized.
This angular momentum, $\ell$, 
can also be regarded as the momentum along the helical axis. 
This leaves us with a differential equation for $\phi(u)$ below:
\begin{equation}
\label{eq:heldir}
\frac{i}{\cosh u}\begin{pmatrix}
0 & {\partial_u}+\ell+{\frac{\tanh u}{2}}\\
{\partial_u}-\ell+{\frac{\tanh u}{2}} & 0
\end{pmatrix}
\begin{pmatrix}
\phi _+ \\
\phi _-\end{pmatrix}
=E\begin{pmatrix}
\phi _+\\
\phi _-
\end{pmatrix}.
\end{equation}
We took a unit in which $a=1$ here. We are going to stick with this unit system
unless otherwise stated.

\subsection{Solving the Equation with fixed $\ell$}
\label{test11}
\begin{figure}[htbp]
\begin{center}
\large{
\includegraphics[width=0.9\columnwidth]{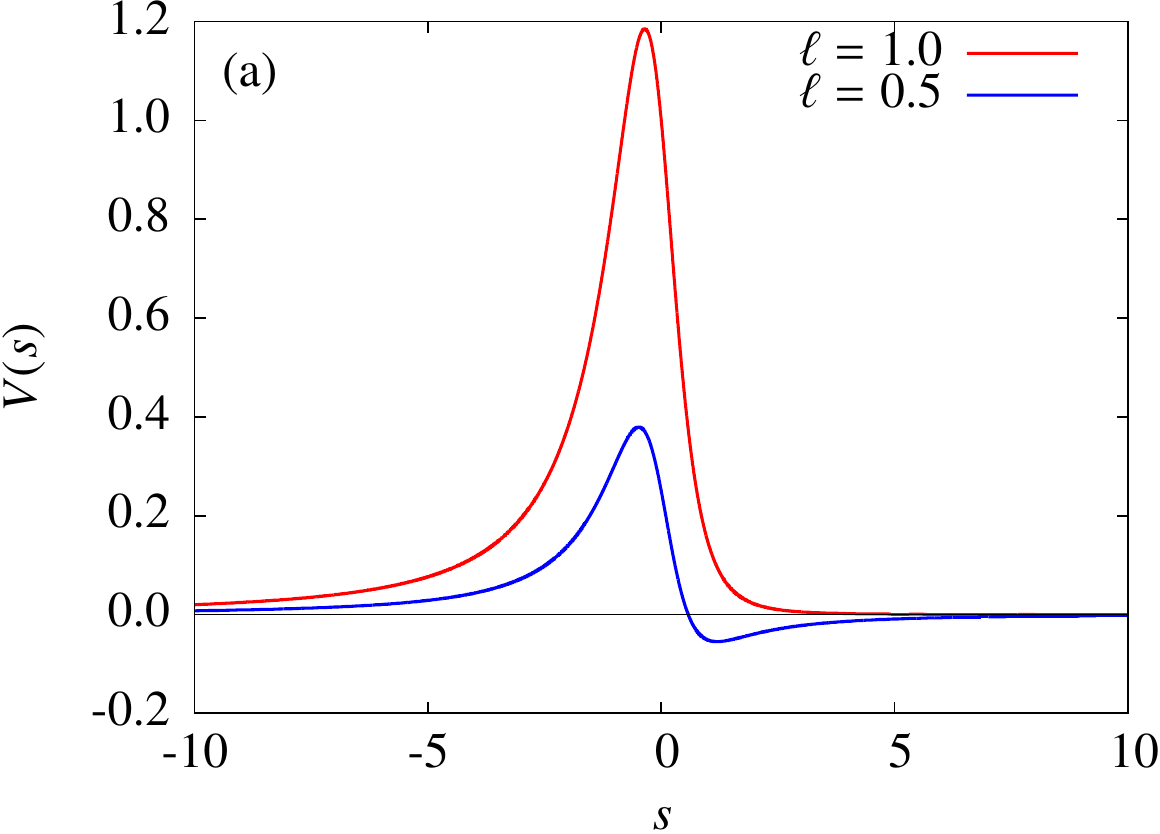}
\includegraphics[width=0.9\columnwidth]{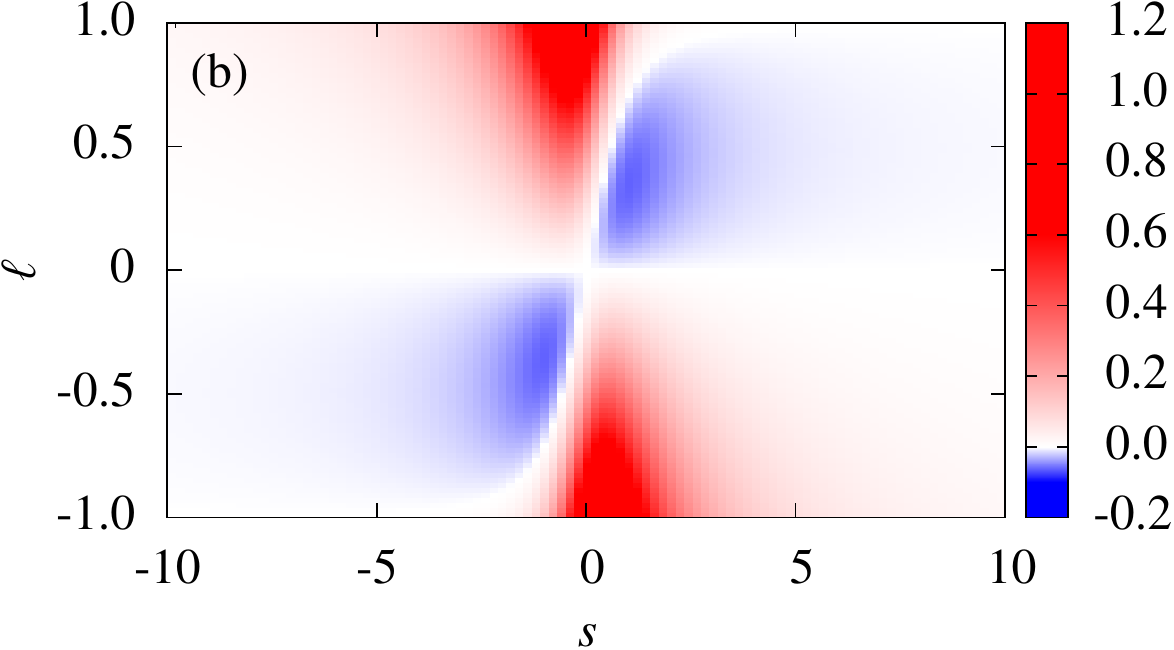}
}
\end{center}
\caption{(Color online) (a) The potential barrier in the 
transformed one-dimensional problem as
 functions of $s=\sinh(u)$ with different values of $\ell$. 
(b) 2D color map of the potential against $\ell$ and $s$.}
\label{potential}
\end{figure}

The massless Dirac equation on the helicoid, (\ref{eq:heldir}), 
is similar to the flat-space Dirac equation in that it takes an
off-diagonal form,
\begin{eqnarray}
\label{eq:heldir2}
&&\begin{pmatrix}
0 & {\frac{iD_{\ell}}{\cosh u}}\\
{\frac{iD_{-\ell}}{\cosh u}} & 0
\end{pmatrix}
\begin{pmatrix}
\phi _+ \\
\phi _-\end{pmatrix}
=E\begin{pmatrix}
\phi _+\\
\phi _-
\end{pmatrix},
\end{eqnarray}
where
\begin{eqnarray}
D_{\ell}=\partial_u+\ell+\frac{\tanh u}{2}.
\end{eqnarray}
We square the Dirac Hamiltonian, 
$
\begin{pmatrix}
0 & {\frac{iD_{\ell}}{\cosh u}}\\
{\frac{iD_{-\ell}}{\cosh u}} & 0
\end{pmatrix}
$,
to get a set of second-order differential equations:
\begin{eqnarray}
\displaystyle\frac{1}{\cosh u}D_{\ell}\frac{1}{\cosh u}D_{-\ell}
\phi_+&=E^2\phi_+
\\
\displaystyle\frac{1}{\cosh u}D_{-\ell}\frac{1}{\cosh u}D_{\ell}\phi_-&=E^2\phi_-
\label{2ndorder}.
\end{eqnarray}
Although it is possible to numerically solve these equations directly,
one alternative is to cast them 
into a
more convenient, Schr\"{o}dinger-type form with the transformations below:
\begin{equation}
s=\sinh(u)\quad \text{and}\quad
\psi_\pm(u)=\phi_\pm(u)\sqrt{\cosh u}
\label{schrtrans}.
\end{equation}
This transformation is chosen for geometric reasons: (i)
$ds$ gives the infinitesimal canonical distance in $\mathbb{R}^3$. (ii)
The integral measure, $\cosh u\,du$, associated with the metric,
Eq. (\ref{metricparam}),
is then factored out.
Following these,
Eq. (\ref{2ndorder}) is transformed into
\begin{equation}
-\frac{d^2}{ds^2}\psi(s)+\left[\frac{\ell^2}{1+s^2}-\frac{\ell s}
{{(1+s^2})^{3/2}}\right]\psi(s)=E^2\psi(s).
\label{1Dschr}
\end{equation}
{
We concentrate
only on the $\psi_-(s)\equiv \psi(s)$ component of the spinor here and hereafter:
we only have to change $\ell$ into $-\ell$ to obtain the $\psi_+(s)$
component of the spinor.
}
Eq. (\ref{1Dschr}), is just the one-dimensional Schr\"{o}dinger equation with 
eigenenergy $E^2$ 
for a particle with mass $m=1/2$ travelling
through the potential barrier
\begin{equation}
V_\ell(s)={\frac{\ell^2}{1+s^2}-\frac{\ell s}{{(1+s^2})^{3/2}}}
\label{potentialbarrier}
\end{equation}
(see Fig. \ref{potential}).
Note that $E^2$ on the right-hand side of Eq. (\ref{1Dschr}) 
ensures we can concentrate only on non-negative 
eigenvalues of this eigenvalue equation. 

\begin{figure}[htbp]
\begin{center}
\large{
\includegraphics[width=0.9\columnwidth]{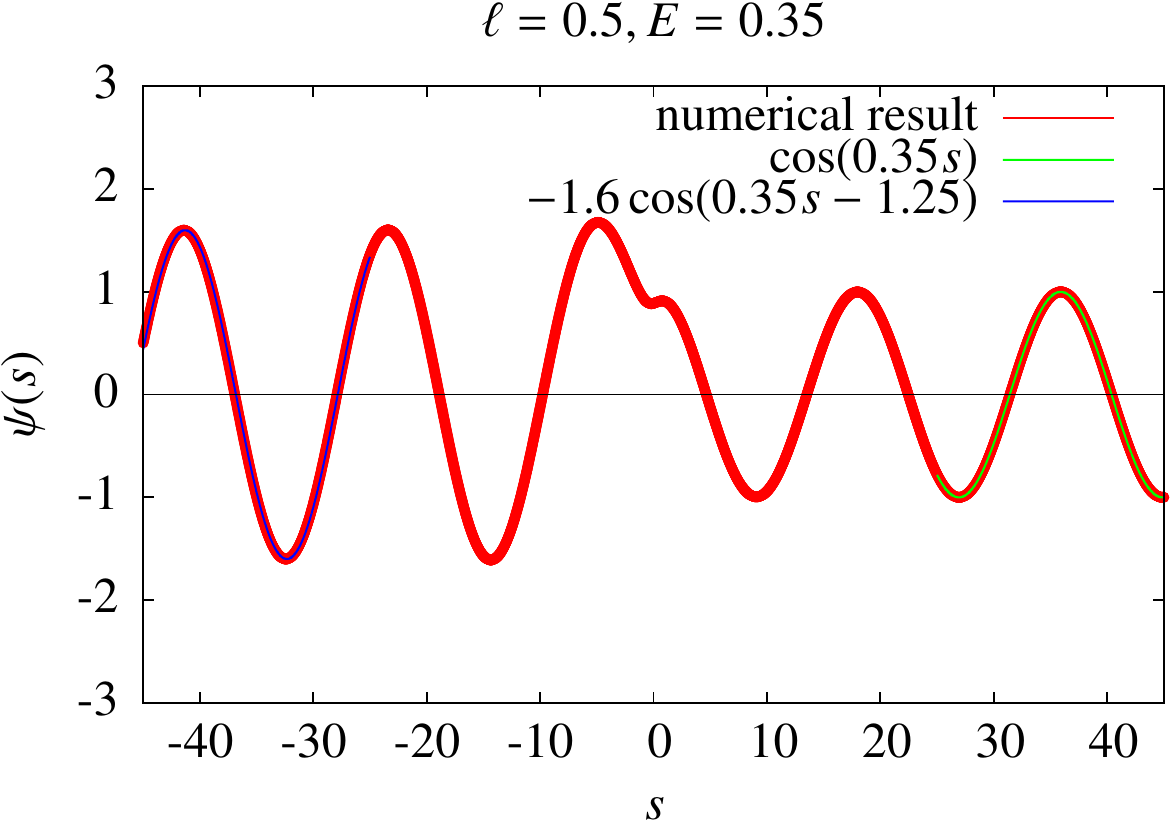}
\includegraphics[width=0.9\columnwidth]{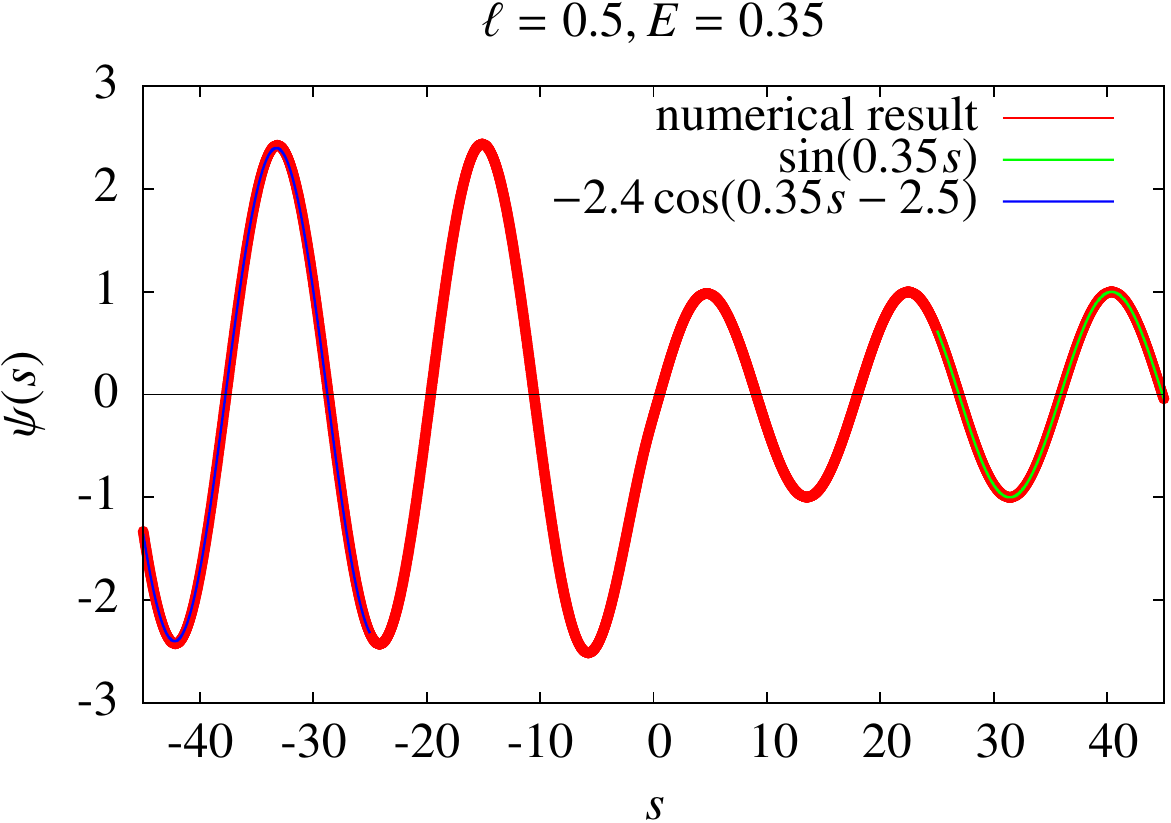}
}
\end{center}
\caption{(Color online) The wavefunction with $\ell =0.5$
at $E=0.35$ (the red, thick curves). 
The green, thin curves on the right side ($25<s<50$)
represent positive asymptotes, i.e.,
$\cos(Es)$ and $\sin(Es)$ in the upper and lower panel, respectively.
The blue, thin curves on the left side ($-50<s<-25$)
are fitted curves representing negative asymptotes.
}
\label{fig:wf1}
\end{figure}

The potential barrier, $V_\ell(s)$, approaches zero 
as $s \rightarrow \pm\infty$ (Fig. \ref{potential}).
This implies we have no bound states with energy above zero.
Therefore all the states we need to consider 
(with eigenenergy 
$E^2$ nonnegative)
are scattering states, 
and we have to solve the equation with boundary conditions
in which eigenstates behave asymptotically like plane waves as $s\to\pm\infty$.
Examples of numerically obtained wavefunctions
are shown in Fig. \ref{fig:wf1}, 
along with their asymptotes.
We have employed the Runge-Kutta method and solved the
differential equation backwards from $s_0=45$,
with its initial value conditions chosen to match
the real/imaginary parts of
$\psi(s_0)=\exp(iEs_0)$ and $\psi^{\prime}(s_0)=iE\exp(iEs_0)$.
We can see from the figure that the wavefunctions deviate from the 
asymptotic plane waves in a region very close to the 
axis of the helicoid ($s \sim 0$). 

\begin{figure}[htbp]
\begin{center}
\includegraphics[width=0.8\columnwidth]{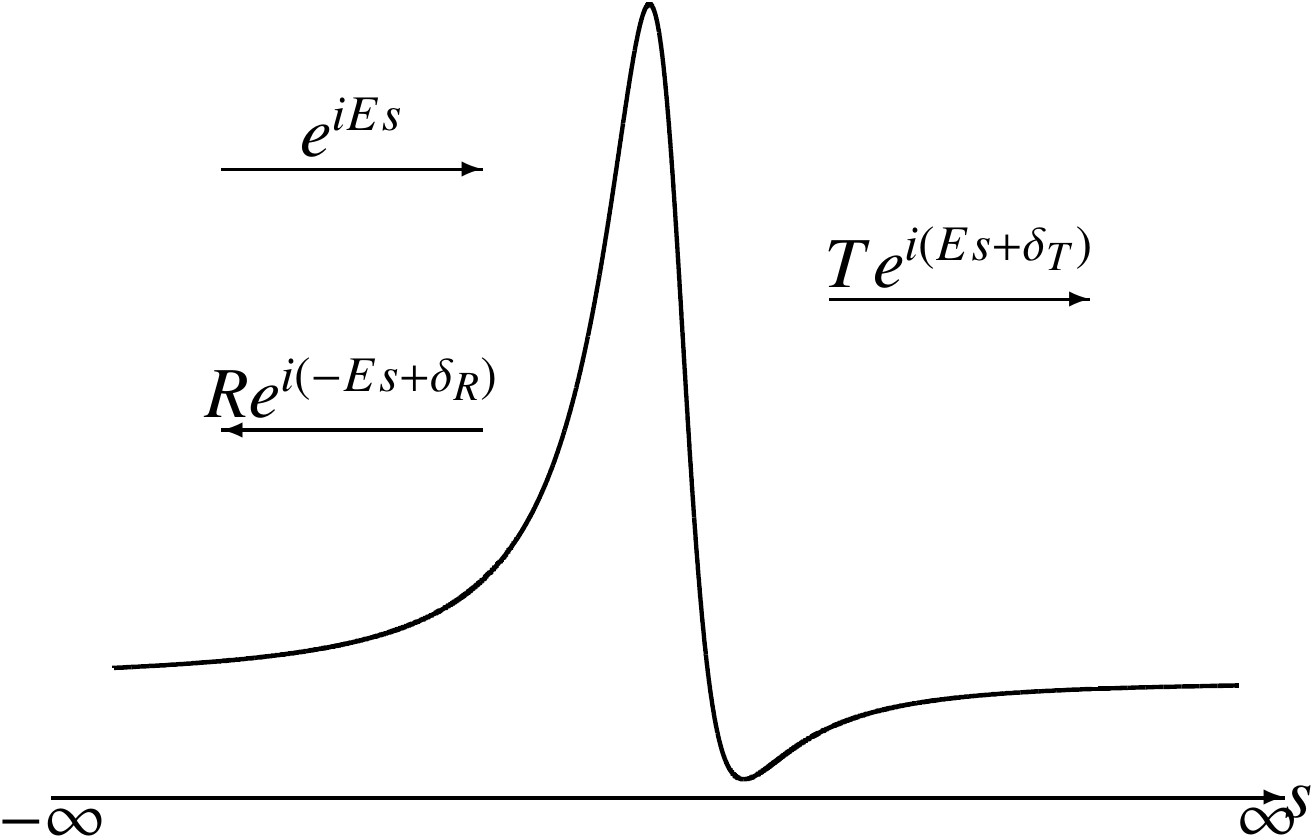}
\end{center}
\caption{The boundary condition of the system as a
transformed one-dimensional problem.}
\label{fig:bc}
\end{figure}

\begin{figure}[htbp]
\begin{center}
\large{
\includegraphics[width=0.9\columnwidth]{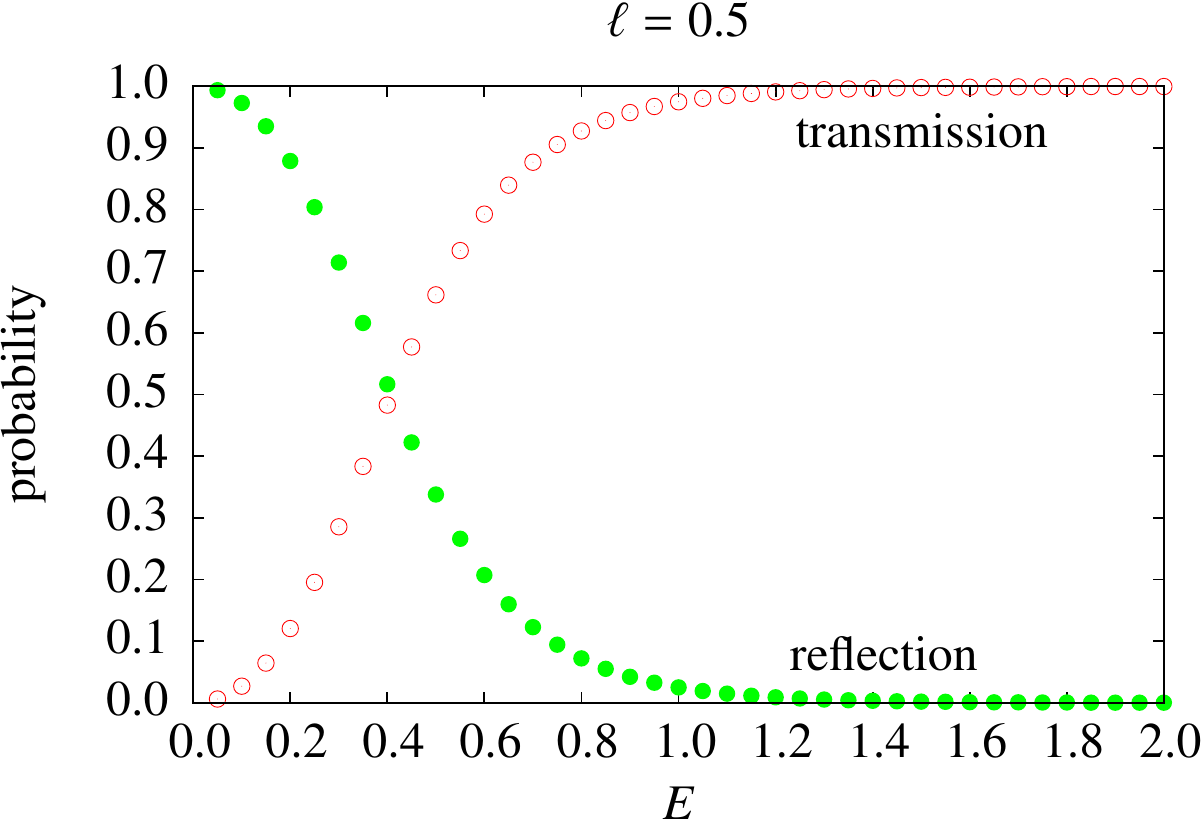}
}

\end{center}
\caption{(Color online) The transmission/reflection probabilities with
$\ell=0.5$
plotted against energy.}
\label{fig:refen}
\end{figure}

We then calculate the reflection and transmission coefficients, 
$\mathcal{R}\equiv Re^{i\delta_R}$ and 
$\mathcal{T}\equiv Te^{i\delta_T}$, using the obtained wavefunctions and
their asymptotes
(see Fig. \ref{fig:bc}). The results are plotted against $E$ in Fig. \ref{fig:refen}.

We can see from the figure that the transmission/reflection probabilities
resemble those of a finite potential barrier, except
that the resonance phenomena does not show up in the present system.
This is because 
 the potential barrier $V_\ell(s)$ is somewhat rounded
compared with finite hard walls.
The wavelength of electrons are much shorter
than the typical scale at which the height of the potential changes.

\subsection{Solving the Equation with fixed $E$}
\begin{figure}[htbp]
\begin{center}
\includegraphics[width=0.9\columnwidth]{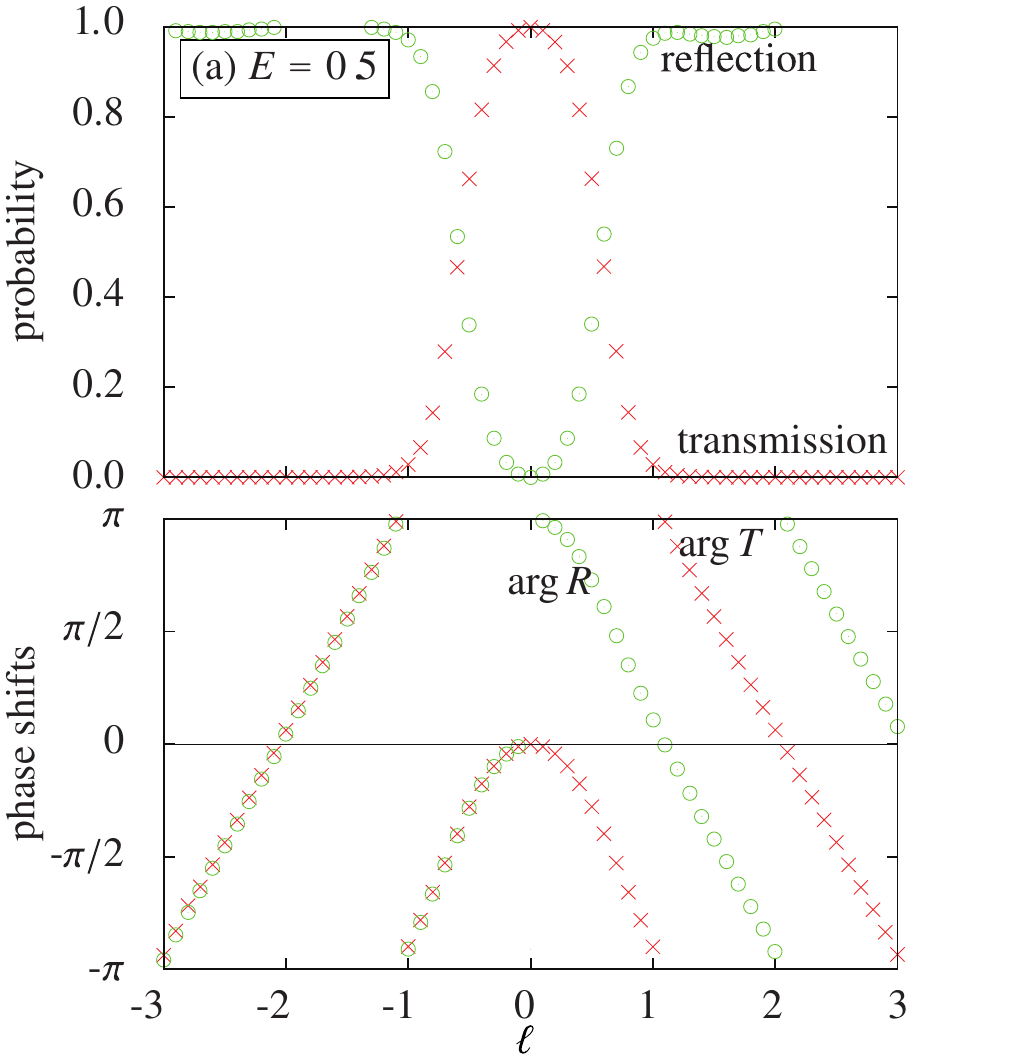}
\includegraphics[width=0.9\columnwidth]{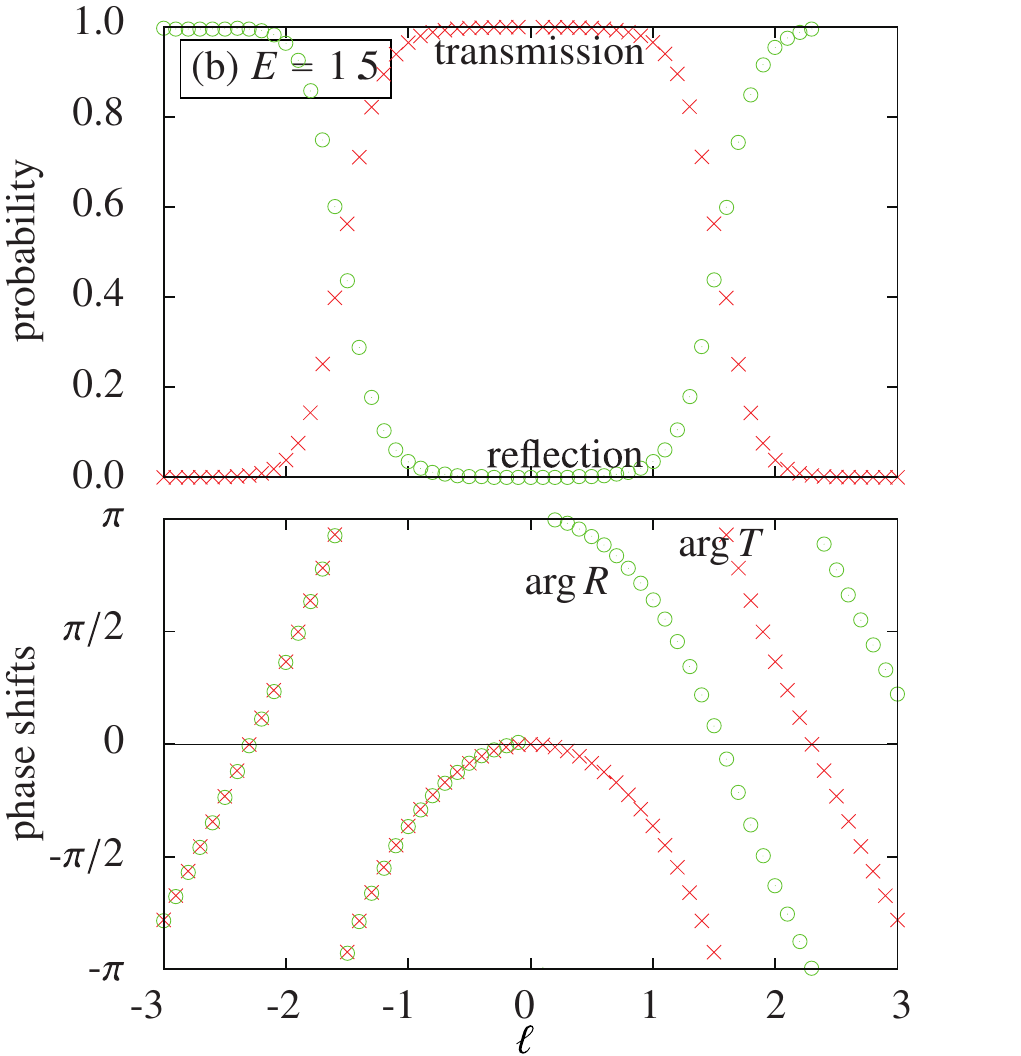}
\end{center}
\caption{(Color online) The transmission and the reflection 
probabilities and the phase shifts plotted against $\ell$ at (a)
$E=0.5$ and (b) $E=1.5$. 
}
\label{fig:t}
\end{figure}

We also plot $R$ and $T$ against $\ell$ (Fig. \ref{fig:t}). 
We can see from Fig. \ref{fig:t}
that $|T|^2$, $|R|^2$ and
$\arg T$
are 
invariant while $\arg R$ increases by $\pi$
under the
inversion of the angular momentum 
($\ell\mapsto -\ell$).
The reason why there is such a symmetry will be
discussed in the next section.

\subsection{Scattering amplitudes and phase shifts against $\ell$ and $E$}
\label{parity}

\begin{figure}[htbp]
\begin{center}
\includegraphics[width=0.9\columnwidth]{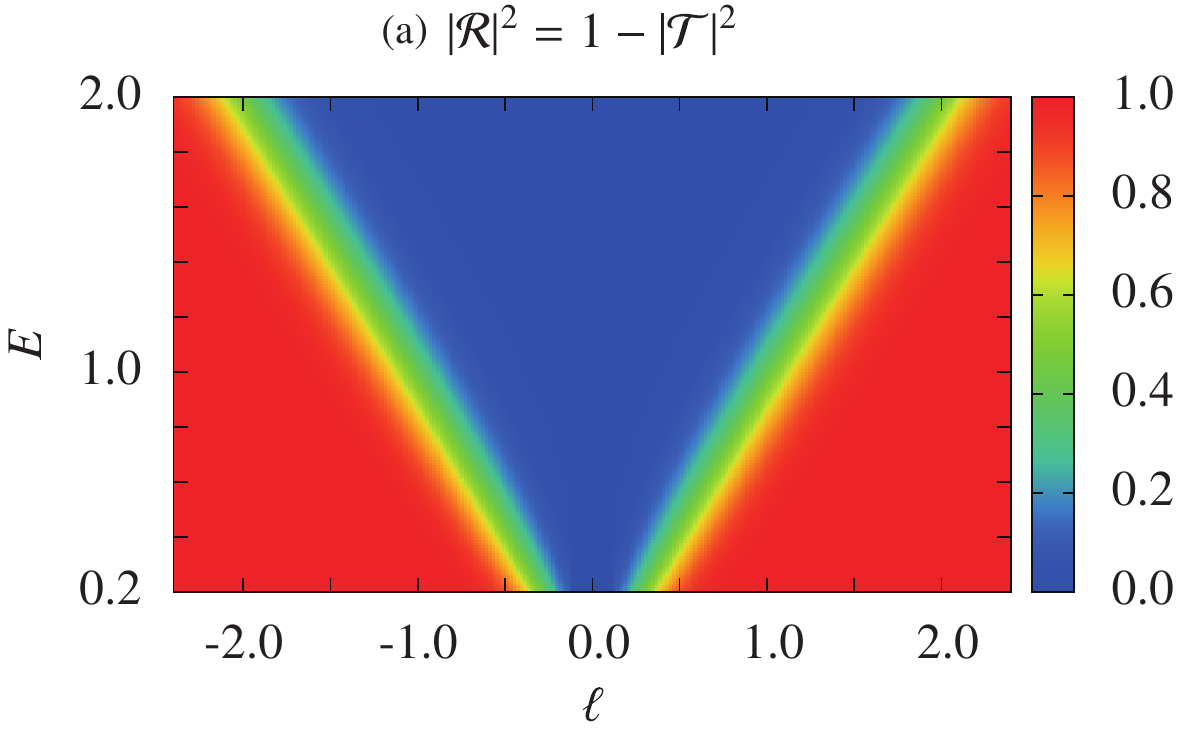}
\includegraphics[width=0.9\columnwidth]{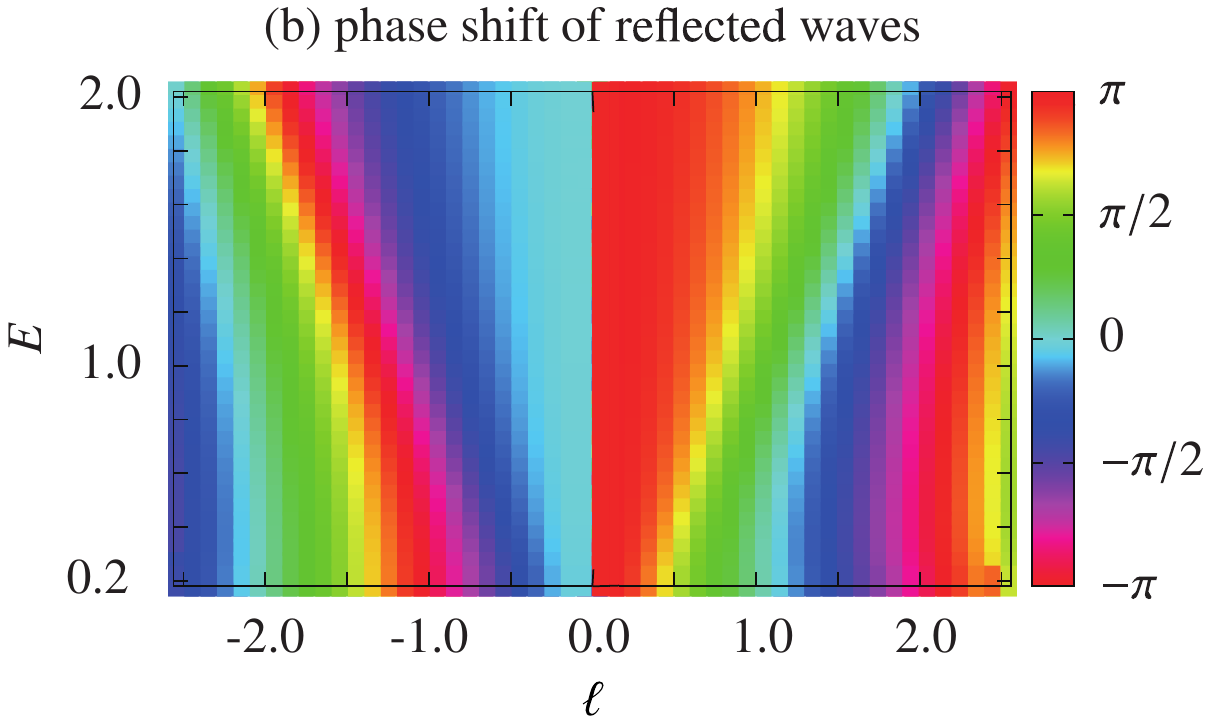}
\includegraphics[width=0.9\columnwidth]{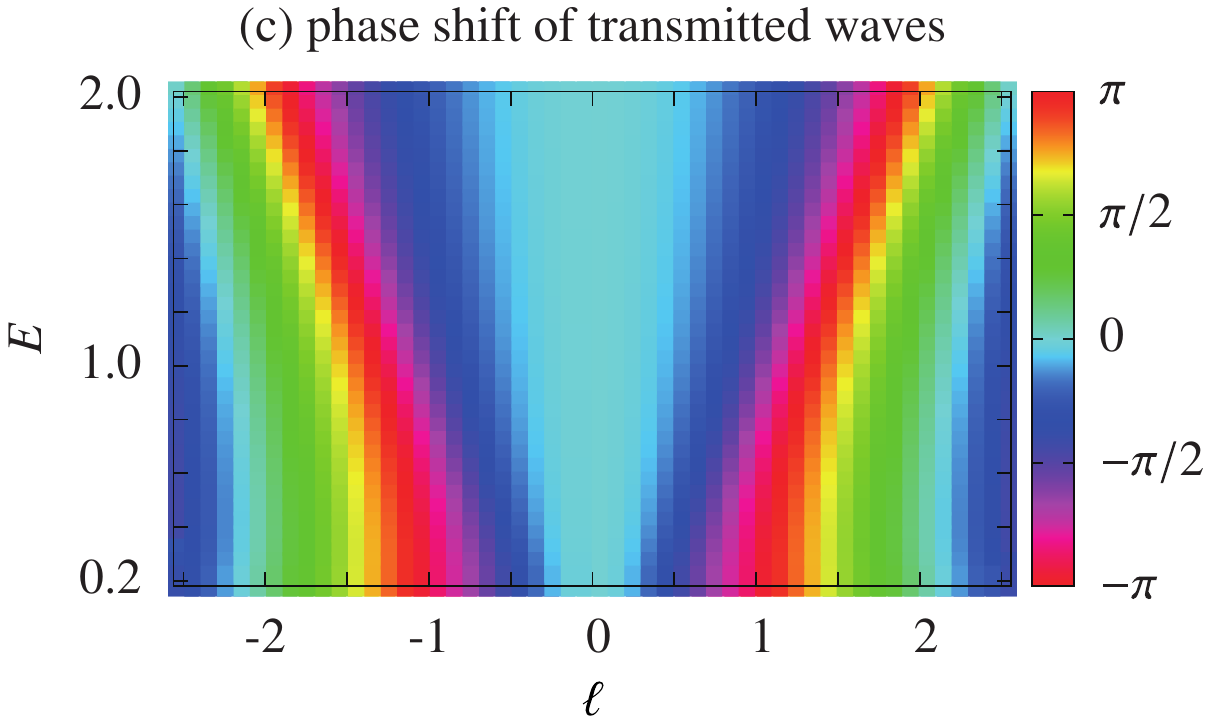}
\end{center}
\caption{(Color online) (a) 2D colour map of the 
reflection probability ($|R|^2 = 1-|T|^2$) 
on $\ell$ and $E$. (b) 2D colour map of the 
reflection phase shifts on $\ell$ and $E$. 
(c) 2D colour map of the transmission phase shifts
on $\ell$ and $E$. 
}
\label{r3d}
\label{phase3d}
\end{figure}

Let us summarize the above results into 
two-dimensional plots against $\ell$ and $E$ of
$|\mathcal{T}|^2$, $|\mathcal{R}|^2$, 
$\arg{\mathcal{T}}$, and $\arg{\mathcal{R}}$. These are
shown in Fig. \ref{r3d}.

Physical quantities tend to change significantly where the eigenenergy $E^2$ becomes
comparable with the peak height of the potential:
\begin{equation}
E^2\sim\frac{2}{27}\left[|\ell|(\ell^2+3)^{3/2}-\ell^2(\ell^2-9)\right]\sim\ell^2\,\,
(\ell\gtrsim O(1)).
\end{equation}
This corresponds to the line where physical
quantities change considerably in Fig. \ref{r3d} (the green regions in (a),
the purple and aqua regions in (b) and the purple regions in (c)).

One interesting question worth asking here is how the 
transmission/reflection coefficients associated with
some positive angular momentum $\ell$
(i.e., electrons spiralling upwards in the $+z$ direction) 
are related to those associated with the inverted angular momentum, $-\ell$
(electrons spiralling downwards in the $-z$ direction). 
In the transformed one-dimensional problem,
the relation $V_\ell(s)=V_{-\ell}(-s)$ implies the following:
inverting the angular momentum ($\ell\mapsto -\ell$)
amounts to inverting $s$ with $\ell$ fixed ($s\mapsto -s$).
We are, therefore,  going to discuss the change in physical quantities
in terms of the space inversion.

\begin{figure}[htbp]
\begin{center}
\includegraphics[width=0.8\columnwidth]{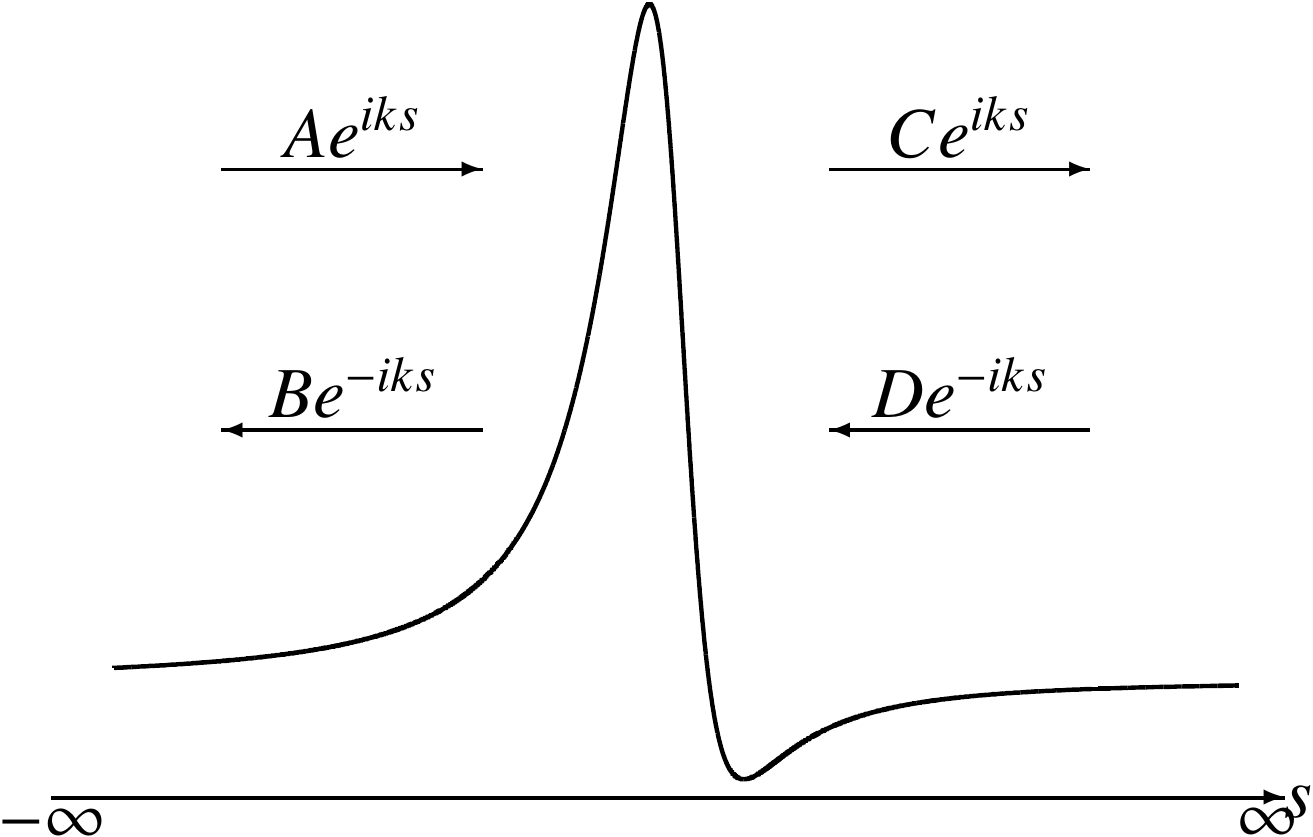}
\end{center}
\caption{Incoming and outgoing waves.}
\label{fig:ten}
\end{figure}

We 
define the transfer matrix as described in Fig. \ref{fig:ten}:
\begin{align}
\label{eq:transm}
\begin{pmatrix}
A\\
B
\end{pmatrix}
=\hat{T}
\begin{pmatrix}
C\\
D
\end{pmatrix},\, \hat{T}=
\begin{pmatrix}
p & r\\
q & s
\end{pmatrix}.
\end{align} 
The relations between the variables above
and those in Fig. \ref{fig:bc} are
$1/p=\mathcal{R}\equiv Re^{i\delta_R}$ and 
$q/p=\mathcal{T}\equiv Te^{i\delta_T}$. 
The probability conservation requires 
$|p|^2-|q|^2=1$ as well as the existence of $\Delta$
such that $r={q^*}e^{i\Delta}$ and $s={p^*}e^{i\Delta}$ hold\footnote{
We can show this as follows: $|A|^2-|B|^2=|C|^2-|D|^2$
leads to $|p|^2-|q|^2=|r|^2-|s|^2=1$ and $p^{*}r-q^{*}s=0$, 
which then leads to
$|p|^2=|s|^2$. 
This means there exists $\Delta$ such that $s=p^{*}e^{i\Delta}$ and
substituting this into the relations above yields $r={q^*}e^{i\Delta}$.
}.

The space-inverted transfer matrix $\hat{T}^\prime$ satisfies, by definition, the following:
\begin{align}
\begin{pmatrix}
D\\
C
\end{pmatrix}
=\hat{T}^\prime
\begin{pmatrix}
B\\
A
\end{pmatrix},\,\,
\hat{T}^\prime=
\begin{pmatrix}
pe^{-i\Delta} & -qe^{-i\Delta}\\
-{q^{*}} & p^{*}
\end{pmatrix}.
\end{align}
This means that the transmission/reflection coefficients,
$\mathcal{T}$ and $\mathcal{R}$, 
of the inverted system are
$e^{i\Delta}/p$ and $-(q^{*}/p)e^{i\Delta}$, respectively. 
The transmission/reflection probabilities, $|\mathcal{T}|^2$ and $|\mathcal{R}|^2$,
are therefore invariant under the 
space inversion. We can also show
\begin{align}
\label{eqeq}
\Delta\approx 0,\,\,
\mathop{\mathrm{arg}} q\propto \frac{1}{E}\to 0\,\, (E\to\infty)
\end{align} 
using the first Born approximation. This is shown in Appendix \ref{app}.
This relation holds to a good approximation, according to our numerical calculation.
This is the reason why, as we can see in Fig. \ref{fig:t},
the phase shift of the transmitted wave is symmetric 
about $\ell=0$ while that of the reflected wave is symmetric otherwise
a jump by $\pi$ at $\ell=0$. 

It might first seem strange that inverting angular momentum 
should bring about changes in any physical quantities.
However, as we are dealing with spinors here\footnote{
Although we started off by dealing with \textit{pseudo} spins on helical graphene,
at this stage we are free to think of these spins as \textit{real} spins; 
there would be no difference in any arguments in the paper according to
whether spins are real or not.
}, 
only when the angular momentum and the spin directions 
are both inverted will the system go back to its original state.
This operation corresponds to inverting $\ell$ and $s$ simultaneously.
\section{Local Density of States}
\label{five}
We now turn to the local density of states (LDoS) of the system.
This is one way of visualising wavefunctions
when dealing with continuous spectra.
LDoS,
$\rho(\mathbf{r},E)$, is defined, in terms of the Green's function, as 
\begin{equation}
\rho(\mathbf{r},E)=\frac{1}{\pi}
\operatorname{Im}\operatorname{Tr}G(\mathbf{r},\mathbf{r},E).
\end{equation}
where $G(\mathbf{r},\mathbf{r}^{\prime},E)$ is the
Green's function and is a $2\times 2$ matrix:
\begin{equation}
{G}(\mathbf{r},\mathbf{r}^{\prime},E)=
\begin{pmatrix}
G^{++}(\mathbf{r},\mathbf{r}^{\prime},E) 
& G^{+-}(\mathbf{r},\mathbf{r}^{\prime},E) \\
G^{-+}(\mathbf{r},\mathbf{r}^{\prime},E) 
& G^{--}(\mathbf{r},\mathbf{r}^{\prime},E) 
\end{pmatrix}
.
\end{equation}
The Green's function for
the Dirac equation
\begin{eqnarray}
\label{ingen}
\begin{pmatrix}
0 & {{D^{+}}}\\
{{D^{-}}} & 0
\end{pmatrix}
\begin{pmatrix}
\psi ^+_n \\
\psi ^-_n\end{pmatrix}
=E_n\begin{pmatrix}
\psi ^+_n\\
\psi ^-_n
\end{pmatrix},
\end{eqnarray}
satisfies the following relation :
\begin{eqnarray}
&&\left[\begin{pmatrix}
0 & {{D^{+}(\mathbf{r})}}\\
{{D^{-}(\mathbf{r})}} & 0
\end{pmatrix}-E\right]
\begin{pmatrix}
G^{++}(\mathbf{r},\mathbf{r}^{\prime},E) 
& G^{+-}(\mathbf{r},\mathbf{r}^{\prime},E) \\
G^{-+}(\mathbf{r},\mathbf{r}^{\prime},E) 
& G^{--}(\mathbf{r},\mathbf{r}^{\prime},E) 
\end{pmatrix}\notag\\
&=&-\delta(\mathbf{r}-\mathbf{r}^{\prime})\mathbf{1},
\end{eqnarray}
This gives a spectral representation of the Green's function:
\begin{eqnarray}
G^{++}(\mathbf{r},\mathbf{r}^{\prime},E)
&=&-\lim_{\delta\to0}
\sum_{n}
\frac{E\psi_n^{+}(\mathbf{r}^{\prime})
\psi_n^{+}(\mathbf{r})}{E^2-E_n^2+i\delta},\\
G^{--}(\mathbf{r},\mathbf{r}^{\prime},E)
&=&-\lim_{\delta\to0}
\sum_{n}
\frac{E\psi_n^{-}(\mathbf{r}^{\prime})
\psi_n^{-}(\mathbf{r})}{E^2-E_n^2+i\delta},
\end{eqnarray}
where we have assumed the completeness of $\{\psi_n^{+}(\mathbf{r})\}$ and of
$\{\psi_n^{-}(\mathbf{r})\}$, respectively. LDoS is then given by
\begin{align}
\rho(\mathbf{r},E)
&=\sum_{n}E\delta(E^2-E_n^2)\left[|\psi_n^+(\mathbf{r})|^2
+|\psi_n^-(\mathbf{r})|^2\right],
\end{align}
or 
\begin{align}
\rho(\mathbf{r},E)=\frac{1}{2}\sum_{\alpha=\pm}
\left[|\psi_E^{\alpha,1}(\mathbf{r})|^2+|\psi_E^{\alpha,2}(\mathbf{r})|^2\right],
\end{align}
if we have a continuous spectrum.
Here we let
$\left\{\psi_E^{\pm ,1}(\mathbf{r}),\,\psi_E^{\pm ,2}(\mathbf{r})\right\}$ 
form an orthonormal basis of the eigenspace 
of ${{D^{\pm}(\mathbf{r})}}
{{D^{\mp}(\mathbf{r})}}$ with eigenvalue $E^2$. 


We can Fourier transform Green's function on the helicoid
with respect to $v$ to decompose LDoS
into partial LDoS, $\rho_\ell(s, E)$:
\begin{align}
\rho(s,E)=\sum_\ell\rho_\ell(s,E)=\sum_\ell\sum_{\alpha=\pm}
\left[|\psi_{\ell,E}^{\alpha,1}(s)|^2+|\psi_{\ell,E}^{\alpha,2}(s)|^2\right].
\end{align}

\begin{figure}[htbp]
\begin{center}
\includegraphics[width=0.9\columnwidth]{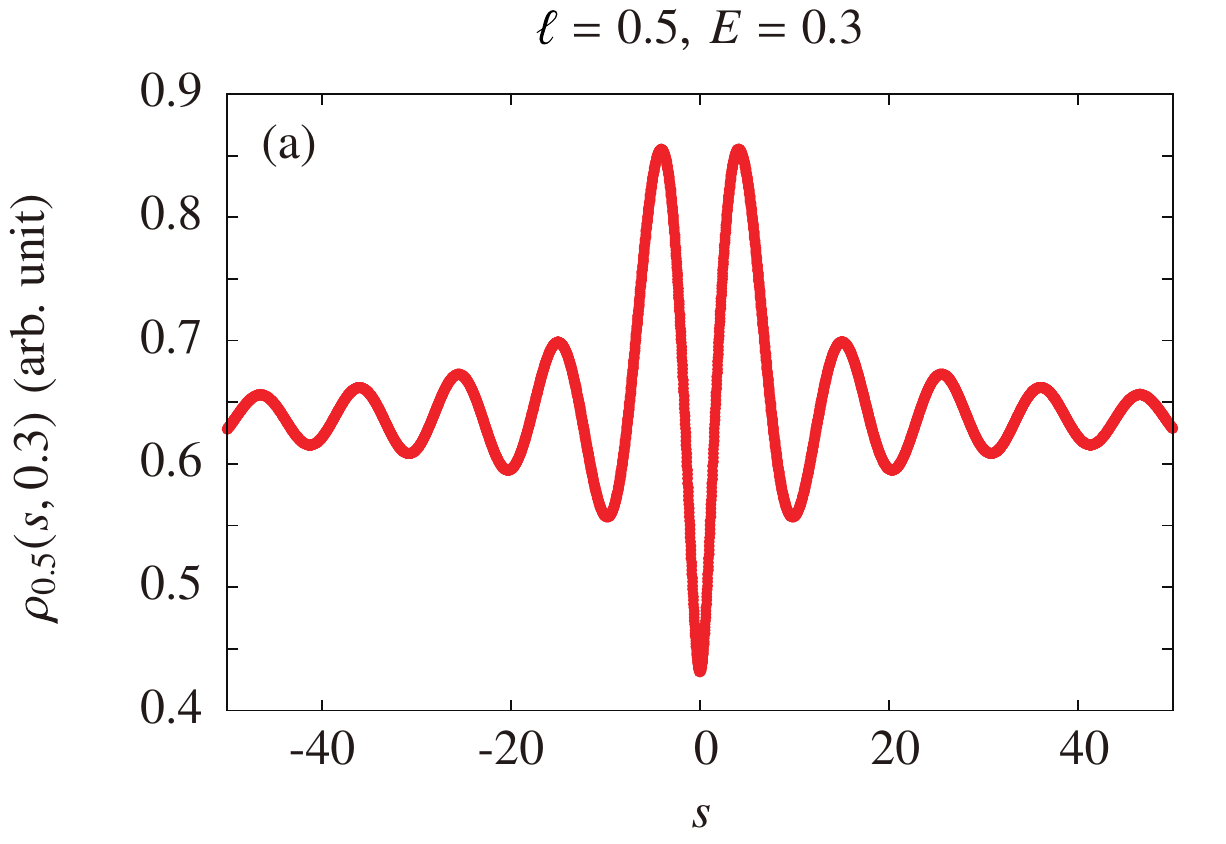}
\includegraphics[width=0.9\columnwidth]{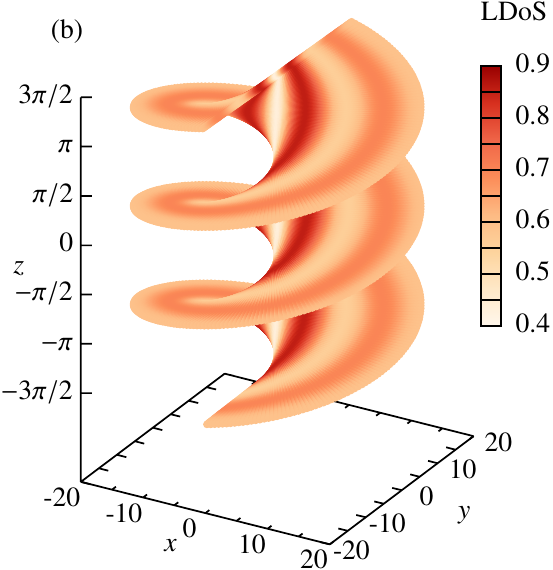}
\end{center}
\caption{(Color online) (a) Local density of states on the helicoid on $s$
with $\ell=0.5$ at
$E=0.3$. Note the symmetry with respect to
the space inversion, $s\mapsto -s$.　
(b) Surface color plot of LDoS on the helicoid.
}
\label{ldos}
\end{figure}

\begin{figure}[htbp]
\begin{center}
\includegraphics[width=0.9\columnwidth]{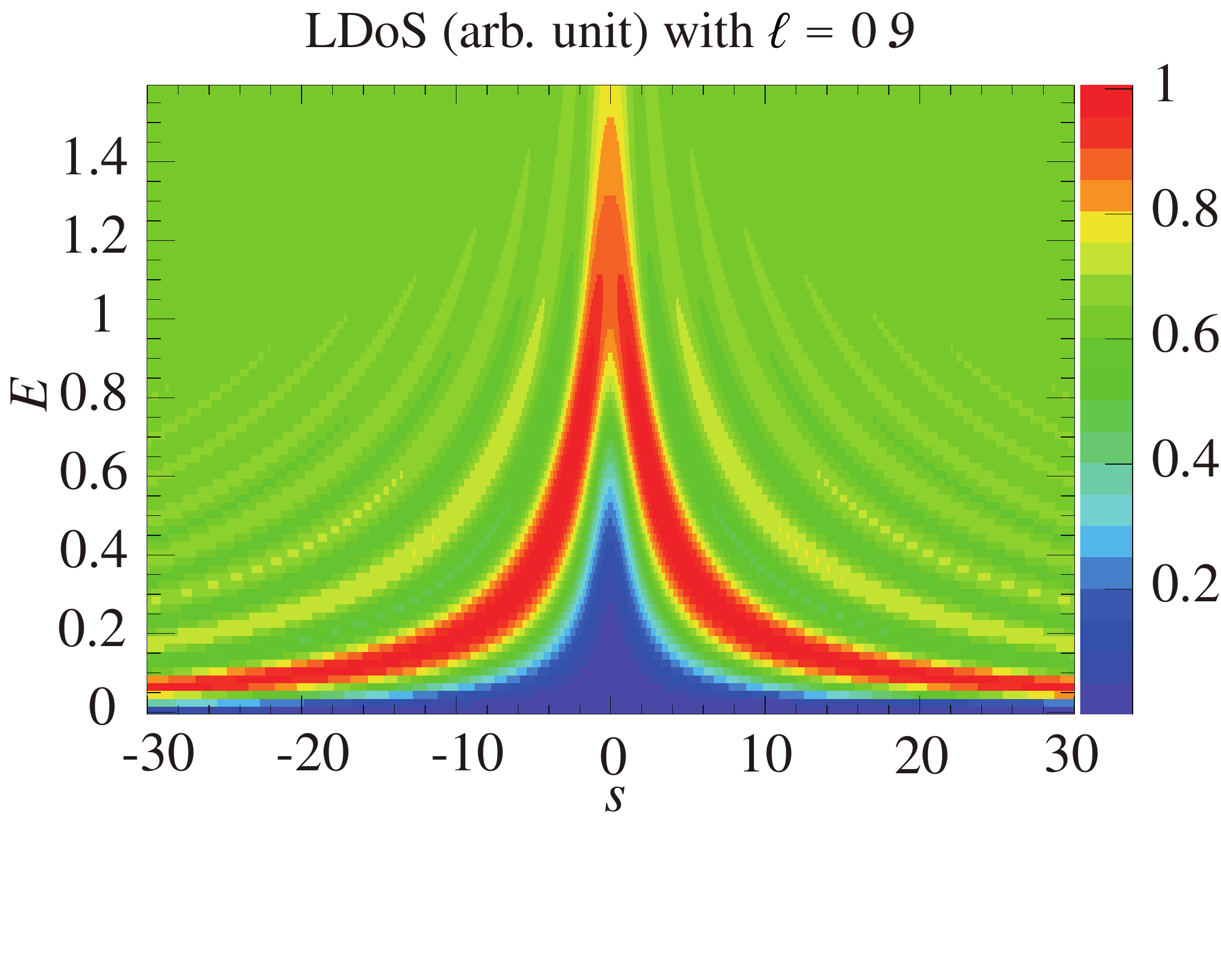}
\includegraphics[width=0.9\columnwidth]{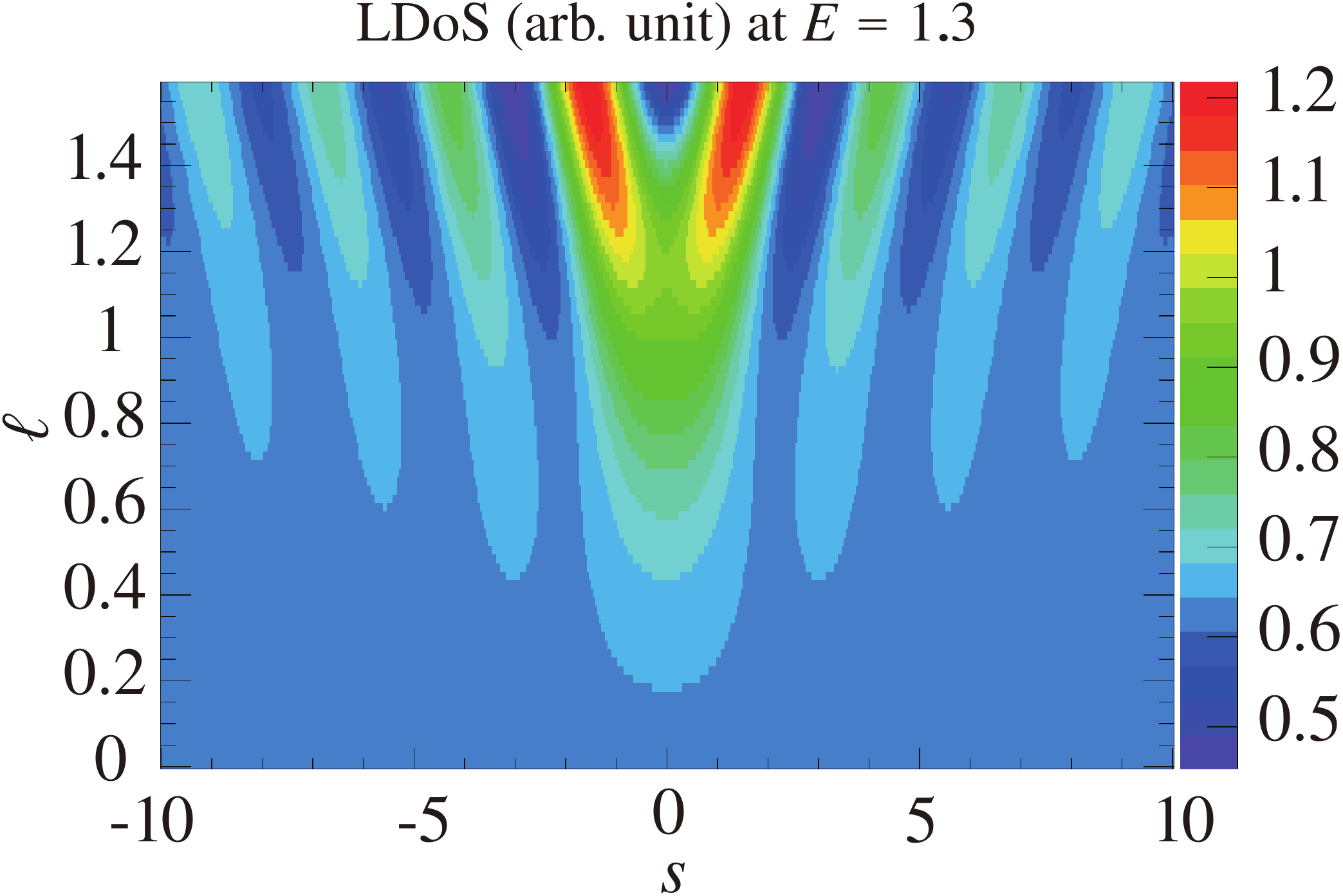}
\end{center}
\caption{(Color online) Local density of states against $E$ and $s$ (upper) 
or against $\ell$ and $s$ (lower).
}
\label{ldos1}
\end{figure}

We calculate partial LDoS numerically.
As the original wavefunctions (asymtoptically $\cos(ks)$ and $\sin(ks)$
 as $s\to\infty$) do not form an orthonormal basis,
we have to prepare $\psi_E^{\pm,1}(s)$ and $\psi_E^{\pm,2}(s)$
which are orthogonal and normalised.
This orthogonalization of the actual wavefunctions
was done 
using the Gram-Schmidt algorithm. 
The numerically calculated LDoS is shown in Fig. \ref{ldos}. 
We can see that LDoS oscillates with large amplitudes in the vicinity of the
axis of the helicoid. 
This is natural because the potential is peaked 
around $s=\sinh(u)\approx 0$ (see Fig. \ref{potential}). 
We also plot LDoS against $E$ and $s$ or against $\ell$ and $s$. These are shown
in Fig. \ref{ldos1}. 
Again LDoS oscillates with large amplitudes around the 
axis of the helicoid in a consistent way:
faster oscillations for larger $E$, and smaller amplitudes
for larger $\ell$.

\section{Massive case}
\label{endd}
Having studied the case of massless Dirac fermions so far,
let us move on to the case of massive Dirac fermions on 
the helicoid. 
One might expect that bound states
could appear in the massive case, for which we replace 
$E^2$ with $E^2-m^2$ in Eq. (\ref{1Dschr}):
\begin{equation}
-\frac{d^2}{ds^2}\psi(s)+\left[\frac{\ell^2}{1+s^2}-\frac{\ell s}
{(1+s^2)^{3/2}}\right]\psi(s)=(E^2-m^2)\psi(s).
\label{1Dschr1}
\end{equation}
We can however prove that bound states continue to be absent
even in the massive case.

In order to show the absence of bound states with
$E^2-m^2<0$, 
we only have to find one wavefunction whose eigenenergy satisfies $E^2-m^2=0$
that is everywhere nodeless. 
This can most easily be done by going back to the original Dirac equation,
(\ref{eq:heldir}). 
At $E=0$, this equation becomes, for $\phi_-$,
\begin{equation}
\left[\frac{d}{du}+\ell+\frac{\tanh(u)}{2}\right]\phi_-=0,
\end{equation}
which can be solved analytically to give 
\begin{equation}
\phi_-(u)=\frac{e^{-\ell u}}{\sqrt{\cosh u}}.
\end{equation}
This can be translated into a wavefunction of the Schr\"{o}dinger-type
equation, (\ref{1Dschr1}), with its eigenenergy satisfying $E^2-m^2=0$:
\begin{equation}
\psi_-(s)=\phi_-(u)\sqrt{\cosh u}=e^{-\ell u}>0,
\label{agg}
\end{equation}
where $s=\sinh u$ as before
\footnote{Change $\ell$ into $-\ell$ to see $\psi_+(s)$ is also positive everywhere.}.
This means that one of the wavefunctions of the states with $E^2-m^2=0$ is nodeless
(hence a ground state wavefunction),
which means there are no bound states at $E^2-m^2<0$.
We therefore conclude the absence of 
bound states even in the case of massive Dirac particles.

We can also show that the state with $E^2-m^2=0$ is actually the
ground state on any surfaces if they have at least one cyclic coordinate.
According to Eq. (\ref{mad}), 
the zero-energy wavefunctions satisfy the following
differential equation:
\begin{equation}
(\partial_x-\partial_y)\left[\sqrt{g(x,y)}\psi_-(x,y)\right]=0
\end{equation}
(see Eqs. (\ref{oha}, \ref{mad}) for how $g$ and $\psi_-$ are defined).
We think of $y$ as the cyclic coordinate
conjugate to the momentum $k$, and the above equation becomes 
\begin{eqnarray}
(\partial_x+k)\left[\sqrt{g(x)}\phi_{-}(x)\right]=0,
\end{eqnarray}
which again can be solved analytically to give 
\begin{eqnarray}
\phi_{-}(x)=\frac{e^{-kx}}{\sqrt{g(x)}}.
\end{eqnarray}
In the Schr\"{o}dinger-type notation the wavefunction is
\begin{equation}
\psi(x)=\phi_-(x){\sqrt{g(x)}}={e^{-kx}}>0
\end{equation}
in agreement with Eq. (\ref{agg}) in the case of the helicoid. Hence the zero-energy 
wavefunction have no nodes, and we 
again identify it with the ground state
wavefunction.

Turning back to the massless case,
the absence of (positive energy) bound states can be loosely accounted for the
Gauss curvature
approaching zero as $u$ approaches positive/negative
infinity.
This is because the effective potential
stemming from the curvature of the surface
is proportional to the Gauss curvature\cite{AokiKoshino}.
We therefore expect
the potential barrier
in the transformed one-dimensional problem, $V_\ell(s)$, to be roughly
proportional to the Gauss curvature of the surface.

\section{Conclusions and Outlook}
We have studied the Dirac electrons on the helicoid so far.
We saw that the scattering mainly occurs just around the helical axis
and that the bound states are absent even when we make fermions massive.

One of the candidates of the physical realization of the present system
is graphite with a screw dislocation.
As our analysis was done 
using the Dirac equation and hence ignoring the actual lattice,
the analysis is applicable when the typical length scale at which
the carbon honeycomb lattice bends is much larger than the lattice constant.

Although we showed the absence of bound states of massless/massive Dirac
electrons on the helicoid,
we can also build surfaces that, unlike helicoid, allow bound states for massive/massless 
Dirac fermions. {
An example of such surfaces is the one whose metric is given by
$ds^2=(dx^2+dy^2)/(y^2+A^2)$ with $A$ being constant.
}
This has a larger curvature at larger distances and the corresponding effective
potential is larger at greater distances.

Further extensions of our work may include applications to
surfaces periodic in two or three 
directions rather than one.
The Dirac equation on those  
surfaces can no longer be reduced to the ordinary Sch\"{o}dinger-type equation
unlike the case of the helicoid here.

We have 
considered only psudo-spins in this paper. Thus
another possible extension of the paper will be to introduce real spins into the analysis:
this will be of great interest with more complicated spin-connections.
This possible research direction is also thought-provoking 
as we might be able to consider spin current on graphene
with combined pseudo- and real spins.

After the submission of the present work, we came to notice 
a paper which deals with helicoidal graphene nanoribbons\cite{PhysRevB.92.035440}.
While
the screw axis region of helicoidal nanoribbons, which
is the subject of the above paper, are hollow,
our main interest in the present paper is the very region, because of its singularity 
and possible applications to graphite with a screw dislocation.

\section{Acknowledgement}
The authors wish to thank Takahiro Morimoto for illuminating 
discussions at the early stages of the present study.  
The present work was supported by Grants-in-Aid for Scientific 
Research (Grant Nos. 25107005, 26247064) from MEXT and ImPACT from JST.

\appendix
\section{Derivation of Eq. (\ref{eqeq})}
\label{app}
We take a notation in which $x=s$ and $k=E$ throughout this appendix. 
The one-dimensional Lippmann-Schwinger equation reads 
\begin{align}
\psi(x)=e^{ikx}-\frac{i}{2k}\int^\infty_{-\infty} dy\,e^{ik|x-y|}V(y)\psi(y).
\end{align}
Approximation \`a la Born gives the first-order perturbation:
\begin{align}
\psi(x)&=e^{ikx}-\frac{i}{2k}\int^\infty_\infty dy\,e^{ik|x-y|}V(y)e^{iky}\notag\\
&=e^{ikx}-\frac{i}{2k}\left[e^{-ikx}\int_x^\infty
 dy\,e^{2iky}V(y)+e^{ikx}\int^x_{-\infty} dy\,V(y)\right].
\end{align}
Hence the first-order perturbative expression for the transmission/reflection coefficients are
\begin{align}
T=1-\frac{i}{2k}\int^\infty_{-\infty}dy\,V(y),\,R=\frac{i}
{2k}\int^\infty_{-\infty}dy\,e^{2iky}V(y).
\end{align}
We can readily see that $T$ is invariant under space inversion\footnote{That $T$ is 
invariant under space inversion can also be seen by 
considering Wronskian and equating its values at $x\to\pm\infty$.} 
and that $\arg T \sim 1/k$ for sufficiently large $k$. Similarly, 
\begin{align}
\int^\infty_{-\infty}dy\,e^{2iky}\frac{1}{1+y^2}&=\pi e^{-2|k|},\nonumber \\
\int^\infty_{-\infty}dy\,e^{2iky}\frac{y}{{\sqrt{1+y^2}}^3}&=4ikK_0(2|k|)
\sim 4ike^{-2|k|}.
\end{align}
so that $\arg R \sim 1/k$ for sufficiently large $k$. This completes the derivation of Eq. 
(\ref{eqeq}).

\section{Nonrelativistic Fermions on the Helicoid}
\label{nonre}
We deal with
with the Schr\"{o}dinger equation on the helicoid 
following the unpublished work by Aoki and Morimoto\cite{Morimoto} cited in the
introduction
for the purpose of comparison with the Dirac case. 

We first write down the Schr\"{o}dinger equation on a curved surface with
metric $g_{ij}$:
\begin{equation}
\left[ 
-\frac{\hbar^2}{2m}\frac{1}{\sqrt{g}} \frac{\partial}{\partial q^i}
\sqrt{g}g^{ij} \frac{\partial}{\partial q^j}
-\frac{\hbar^2}{8m}(\kappa_1 -\kappa_2)^2
\right]
\phi(q^k) 
=E\phi(q^k),
\end{equation}
where $(q^1,q^2)$ is the coordinate system on the surface, denoted hereafter as $(u, v)$,
and $\kappa^1$ and $\kappa^2$ are the two principal curvatures on the surface. 

Specifically, the Schr\"{o}dinger equation on the helicoid (metric given by
 (\ref{metricparam}))
becomes
\begin{equation}
\frac{1}{\cosh^2u}\left(\frac{d^2}{du^2}+\ell^2-\frac{1}
{\cosh^2u}\right)\phi(u)=E\phi(u).
\end{equation}
We used the fact that the two principal curvatures of the helicoid is
$\kappa_{\pm}=\pm 1/(a\cosh^2(u))$\cite{aw}.
We also set $a=1$, $\hbar=1$ and $m=1/2$, as in the case of the Dirac particles.
Here, rather surprisingly, the coordinate transformation for 
the Dirac equation, Eq. (\ref{schrtrans})
can also be applied to transform this Schr\"{o}dinger equation into
a new Schr\"{o}dinger equation on the flat metric:
\begin{equation}
-\frac{d^2}{ds^2}\psi(s)+\left(\frac{\ell^2}{1+s^2}-\frac{s^2+2}{4(1+s^2)^2}
\right)\psi(s)=E\psi(s).
\end{equation}
The potential of this Schr\"{o}dinger equation
is $V(s)=\frac{\ell^2}{1+s^2}-\frac{s^2+2}
{4(1+s^2)^2}$, which is different from that of the Dirac case.

We first note that the system does not allow for bound states,
following a similar argument in the main body of the text. The absence
of bound states on the helicoid, therefore, is just not specific to
Dirac fermions. 

We also see that the potential is symmetric with respect to $\ell\mapsto-\ell$ and
$s\mapsto -s$.
This indicates that the reflection phase shift does not change by $\pi$
at $\ell=0$. This is one major difference in physical quantities
between the Dirac and the Schr\"{o}dinger case. As mentioned at 
the end of Section \ref{parity}, the symmetry 
of the Dirac system is the 
combined inversion of the angular momentum and the (pseudo-)spin
directions, whereas the allowed symmetry operation in the Schr\"{o}dinger case 
is to invert the angular momentum only -- no spins are involved in the Schr\"{o}dinger
system. 

\begin{figure}[htbp]
\begin{center}
\includegraphics[width=0.9\columnwidth]{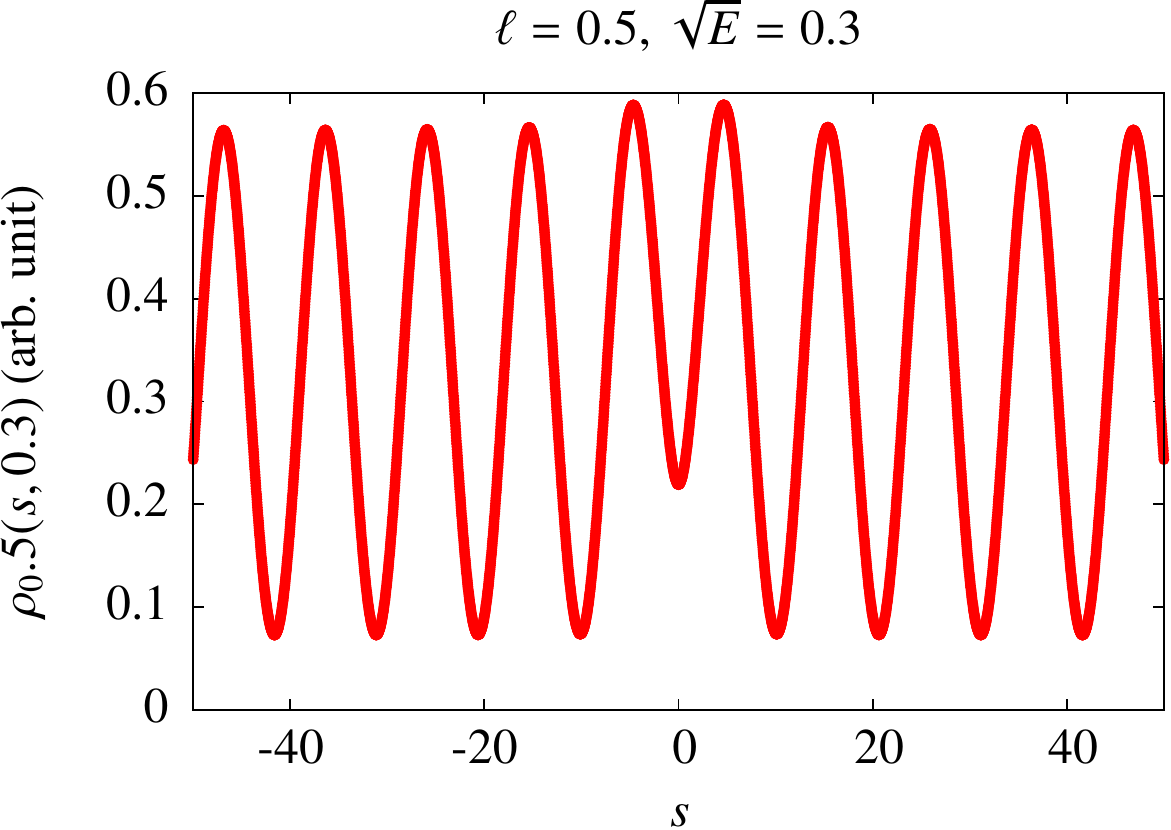}
\end{center}
\caption{(Color online) Local density of states 
in the Schr\"{o}dinger case with $\ell=0.5$ at
$\sqrt{E}=0.3$.
}
\label{dossch}
\end{figure}

LDoS of the Schr\"{o}dinger system, shown
in  Fig. \ref{dossch}, behaves 
also differently from that of the Dirac system. 
The oscillation amplitude of LDoS does not 
diminish as we go farther away from the helical axis
unlike the Dirac case. 

\bibliography{cite}
\end{document}